\documentclass[10pt,conference]{IEEEtran}

\usepackage{cite}
\usepackage{amsmath,amssymb,amsfonts}
\usepackage{algorithmic}
\usepackage{graphicx}
\usepackage{textcomp}
\usepackage{xcolor}
\usepackage{booktabs}
\usepackage[utf8]{inputenc}
\usepackage{xurl} 
\usepackage{etoolbox}
\usepackage{enumitem}
\usepackage{lettrine}
\usepackage{fancyvrb}
\usepackage{tcolorbox}
\usepackage{listings}
\usepackage{multirow}
\usepackage[switch]{lineno} %
\usepackage[strict]{changepage}
\usepackage{framed}

\definecolor{formalshade}{rgb}{0.95,0.95,1}
\definecolor{darkblue}{rgb}{0.0, 0.0, 0.55}
\newenvironment{quotebox}{%
  \MakeFramed{\advance\hsize-\width\FrameRestore}%
  \noindent\hspace{-4.55pt}%
  \begin{adjustwidth}{}{7pt}%
  \vspace{2pt}\vspace{2pt}%
}
{%
  \vspace{2pt}\end{adjustwidth}\endMakeFramed%
}
\usepackage{hyperref}

\AtBeginEnvironment{quote}
{\par\singlespacing\small}

\colorlet{lightergray}{lightgray!50!white}

\def\BibTeX{{\rm B\kern-.05em{\sc i\kern-.025em b}\kern-.08em
    T\kern-.1667em\lower.7ex\hbox{E}\kern-.125emX}}

\begin{document}

\newcommand{\rqi}{What are the characteristics of PTMs used in software projects?}
\newcommand{\rqii}{How are PTMs used in software projects?}
\newcommand{\rqiii}{How do PTMs evolve in the software projects?}
\newcommand{\rqiv}{What are the testing practices related to PTMs in software projects?}
\newcommand{\rqv}{What insights can be gained from issues related to PTMs in software projects?}

\newcommand{\ts}{\textsuperscript}

\newcommand{\diego}[1]{\textcolor{red}{Diego: #1}}
\newcommand{\tocite}{\textcolor{red}{[]}}

\title{Exploring the Lifecycle and Maintenance Practices of Pre-Trained Models in Open-Source Software Repositories }

\author{
    \IEEEauthorblockN{Matin Koohjani, Diego Elias Costa} \IEEEauthorblockA{Department of Computer Science and Software Engineering \\ Concordia University \\ Montreal, Quebec, Canada \\
    matin.koohjani@mail.concordia.ca, diego.costa@concordia.ca}
}

\maketitle

\begin{abstract}
Pre-trained models (PTMs) are becoming a common component in open-source software (OSS) development, yet their roles, maintenance practices, and lifecycle challenges remain underexplored. This report presents a plan for an exploratory study to investigate how PTMs are utilized, maintained, and tested in OSS projects, focusing on models hosted on platforms like Hugging Face and PyTorch Hub. We plan to explore how PTMs are used in open-source software projects and their related maintenance practices, by mining software repositories that use PTMs, and analyze their code-base, historical data, and reported issues.
This study aims to provide actionable insights into improving the use and sustainability of PTM in open-source projects and a step towards a foundation for advancing software engineering practices in the context of model dependencies.

\end{abstract}

\section{Introduction}

\lettrine{T}{he} rapid development and adoption of pre-trained models (PTMs) have transformed the field of artificial intelligence (AI), particularly in areas such as natural language processing and computer vision. Platforms such as Hugging Face and PyTorch Hub have emerged as vital hubs for sharing and distributing these models, offering a wide catalogue of PTMs that developers and researchers can integrate into their projects without requiring extensive resources to train models from scratch. This accessibility has significantly lowered the barrier to entry for deploying state-of-the-art AI capabilities, enabling a growing number of open-source software (OSS) projects to leverage PTMs effectively.

However, while PTMs offer immense benefits, they also introduce unique challenges for OSS projects. Unlike traditional software libraries, PTMs often lack standardized practices such as semantic versioning, which complicates the tracking of updates, compatibility assessments, and lifecycle management. This lack of transparency creates potential risks for developers, who may inadvertently rely on outdated or unsupported models, jeopardizing the stability and reliability of their software. Additionally, the roles these models play within projects vary widely, from serving as core components to being used in experimental or illustrative contexts, further highlighting the need to understand their integration and usage patterns.

Building on the PeaTMOSS dataset~\cite{Jiang:10.1145/3643991.3644907}, which documents thousands of OSS repositories using PTMs from platforms like Hugging Face and PyTorch Hub, this study seeks to explore critical questions surrounding the use and maintenance of PTMs in OSS. Specifically, we aim to investigate how these models are utilized in real-world Python projects, focusing on the model characteristics (RQ1), their roles within the project (RQ2), their evolution and update practices (RQ3), their testing (RQ4), and their potential challenges (RQ5). By exploring these aspects, we hope to shed light on the challenges the OSS community faces and provide insights into improving practices for integrating and maintaining PTMs in software development.
In this study, we propose five research questions that guide our investigation:

\begin{itemize}
    \item[\textbf{RQ1:}] \textbf{\rqi} \\
    In this RQ, we will build a profile of the PTMs being integrated into the most popular OSS projects. We plan to collect and analyze metadata about these models, including licensing information, model type (such as text or image), size, and architecture. The Hugging Face model hub will be our primary source for this information, but we anticipate challenges in obtaining complete metadata for every model we encounter.
    
    \item[\textbf{RQ2:}] \textbf{\rqii} \\
    We plan to investigate the role of PTMs in their respective software projects and how developers incorporate them into the code. 
    To evaluate the role of PTM models, we rely on quantifying their centrality in the software project and manually assess their context of usage (e.g., core functionality, testing, proof-of-concept).   
    Finally, we plan to understand the coding practices for loading PTMs to understand how PTMs are integrated and used within different software projects.

    \item[\textbf{RQ3:}] \textbf{\rqiii} \\
    In this question, we dive into a longitudinal analysis to understand the maintenance and evolution of PTMs in the OSS projects. 
    We examine how many PTMs are included over time, and for how long they remain active in the code (i.e., their lifecycle). 
    Additionally, we plan to evaluate how frequently PTMs are updated to account for the latest versions of PTMs.

    \item[\textbf{RQ4:}] \textbf{\rqiv} \\
    We plan to explore if and how developers test their PTM components in their software projects. This involves analyzing the general testing practices of a project, evaluating test coverage specifically for the code related to PTMs, and identifying any test cases designed explicitly to test PTMs' behaviour. 

    \item[\textbf{RQ5:}] \textbf{\rqv} \\
    We will examine project issue trackers to glean insights into the challenges, bugs, and discussions related to PTM usage. This involves developing a set of keywords to identify PTM-related issues effectively and then analyzing a stratified sample of these issues.
\end{itemize}
With these research questions defined, our primary motivation is to better understand how PTMs are integrated, utilized, and maintained in real-world software projects. Given the rapid adoption and significant impact of PTMs, there is an urgent need for empirical insights into current practices, trends, and challenges. Our study aims to address this need and provide practical guidance for developers and researchers. The replication package of this exploratory study will be publicly available.

\section{Related Work}
\label{sec:related_work}

\subsection{\textbf{Pre-Trained Models and Software Engineering}}
Pre-trained models (PTMs) have become pivotal in addressing challenges in software engineering, providing reusable solutions to complex tasks. As highlighted by Niu et al. \cite{niu2022deeplearningmeetssoftware}, the application of PTMs for source code tasks has unlocked new possibilities in areas such as defect detection, code summarization, and clone identification. They describe how these models, like CodeBERT and CodeGPT, build upon the success of natural language processing PTMs by adapting to the unique characteristics of source code, such as syntax and semantic structures. Their survey emphasizes the importance of leveraging code-specific adaptations to improve task performance and outlines future directions for enhancing these models in software engineering.

Ajibode et al. \cite{ajibode2024semanticversioningopenpretrained} focus on a fundamental challenge with PTMs: semantic versioning. Their analysis of over 52,000 PTMs on Hugging Face reveals inconsistencies in naming conventions and gaps in documentation. They provide evidence that many PTM releases lack proper indicators of changes, with 40.87\% of weight file updates failing to reflect these modifications in their naming schemes. This study brings attention to the need for standardized practices to ensure that PTMs can be effectively integrated and maintained within software systems.

In their work, Jiang et al. \cite{Jiang:10.1145/3643991.3644907} introduce the PeaTMOSS dataset, which represents a major step forward in documenting the PTM lifecycle. They explain that this dataset contains metadata for over 281,000 PTMs, along with detailed mappings to downstream GitHub repositories. By analyzing these relationships, the authors uncover trends in model reuse, identify common issues in licensing, and provide foundational data for future research on PTM dependencies. This paper emphasizes the critical role of structured metadata in enabling robust PTM adoption and reuse.

The community dynamics surrounding PTMs are explored by Castaño et al. \cite{castaño2024analyzingevolutionmaintenanceml}, who analyze over 380,000 models on Hugging Face. They note that the platform has become a hub for collaborative development, enabling researchers and practitioners to share and refine ML models. By studying the evolution of model cards and tags, they uncover patterns in documentation and metadata practices, which are crucial for ensuring transparency and reproducibility in PTM-based projects.

\subsection{\textbf{Maintenance and Evolution of Software Projects Using Pre-Trained Models}}
Maintaining and evolving software projects that rely on PTMs is a complex and dynamic task. Latendresse et al. \cite{Latendresse:10.1145/3688841} take a deep dive into these challenges, proposing a taxonomy of machine learning model management activities based on an analysis of 227 GitHub repositories. They find that over 57\% of these activities are related to maintenance, with tasks such as refactoring and documentation taking center stage. Their findings highlight the importance of automation tools to streamline these processes and address the growing complexity of managing PTM-dependent software.

Building on this, Castaño et al. \cite{castaño2024analyzingevolutionmaintenanceml} examine the maintenance practices of models hosted on Hugging Face, categorizing commits into corrective, perfective, and adaptive types. They discuss how these patterns reflect the evolving nature of ML models and underscore the necessity of robust maintenance frameworks to tackle issues like concept drift and versioning inconsistencies. This study provides actionable insights for improving the sustainability of ML models in real-world applications.

The practical challenges of using PTMs are explored by Tan et al. \cite{tan2024challengesusingpretrainedmodels}, who analyze 5,896 questions from Stack Overflow. They identify recurring issues such as fine-tuning, output customization, and memory management, emphasizing the gaps between existing PTM capabilities and practitioners’ needs. Their study offers a roadmap for improving PTM utilization through better tooling, education, and community support, making it a valuable resource for both researchers and practitioners.

The PeaTMOSS dataset, as introduced by Jiang et al. \cite{Jiang:10.1145/3643991.3644907}, plays a crucial role in addressing these challenges by providing detailed mappings between PTMs and their downstream applications. By documenting 44,337 dependencies, this dataset offers a comprehensive view of the PTM ecosystem, enabling researchers to study trends and identify best practices for maintenance and evolution.

Marchezan et al. \cite{marchezan2024modelbasedmaintenanceevolutiongenai} further expand on the role of GenAI in addressing these challenges, arguing that automation and augmentation can reduce the cognitive load on engineers. They propose a classification scheme that outlines how GenAI can assist, reason, and automate various MBM\&E (Model-Based Maintenance \& Evolution) tasks, paving the way for more scalable and sustainable software maintenance practices.

\section{Study Design}
\label{sec:study_design}
\begin{figure*}[]
\centering
\includegraphics[width=\textwidth]{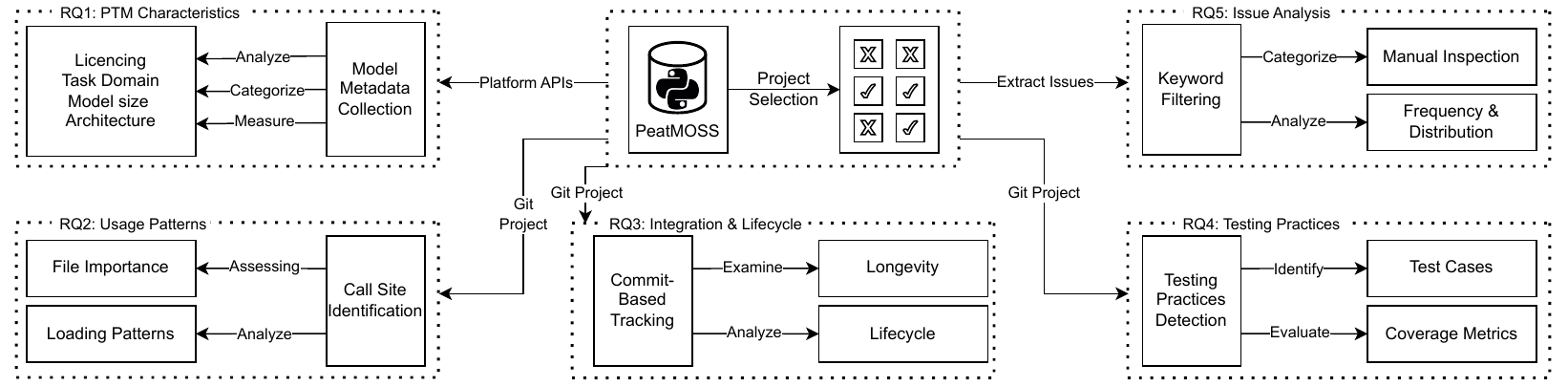}
\caption{Overview of our study}
\label{fig:methodology}
\end{figure*}

This study investigates how Pre-Trained models are used and maintained in open-source GitHub repositories. We build upon the PeaTMOSS dataset, which documents a large collection of Pre-Trained models and their usage in open-source projects. We focus on repositories that are more likely to reflect production-level software. This study aims to address several critical aspects of how developers use, maintain, and update these models over time.

\subsection{\textbf{Data Selection}}

\noindent
\textbf{Candidate Project Selection.}
We build on a subset of the PeaTMOSS dataset, which includes thousands of open-source repositories utilizing Pre-Trained models hosted on Hugging Face and PyTorch Hub. 
The PeaTMOSS dataset was created by a large-scale mining effort to capture OSS projects that use PTMs from Hugging Face and Pytorch Hub~\cite{Jiang:10.1145/3643991.3644907}.
In total, the PeaTMOSS dataset has 44,337 mappings from 15,219 OSS projects to the 2,530 PTMs they use.   
To contextualize, the set of projects was identified by their use of Python library APIs that load PTMs from both Hugging Face and Pytorch Hub. 
Hence, the set of projects are written entirely in Python or contain Python modules to load the PTMs. 
To ensure our analysis focuses on active and significant projects, we apply the following filters:
\begin{itemize}
    \item Repositories where their codebase is written primarily in Python, to unify our methodology when using static analysis tools and testing frameworks. 
    \item Repositories with more than 200 stars.
    \item Projects with at least 10 contributors and at least 10 reported issues, to ensure sufficient data for issue analysis.
    \item Project contains at least 10 commits recorded in 2024 or later, to ensure a healthy level of code activity.
    
\end{itemize}
Although the numerical thresholds selected for project filtering (e.g., minimum stars, contributors, and commits) are not absolute, they are well-established and commonly adopted criteria in empirical software engineering. These thresholds effectively ensure the inclusion of active, influential, and representative repositories, increasing the validity and relevance of our study’s findings, as demonstrated by similar practices in prior studies~\cite{kalliamvakou2014promises, jafari-npm, Costa:GO:2022, Ozren:2021:SamplingMSR}
After applying these criteria, we select a total of 790 repositories for candidates for our analysis. 
We include the descriptive statistics of the projects in Table~\ref{tab:project_stats}.
These repositories are representative of popular, active software projects, where half the projects (shown in the median column) have at least 832 commits, are developed by 29+ contributors, and are at least 3 years old. 

\noindent
\textbf{Excluding educational projects.}
We plan to exclude popular educational projects from our dataset, as they do not represent the subjects we aim to study. 
To achieve this, we will perform a keyword filtering step on the project title and description, similar to other previous studies~\cite{Costa:GO:2022}. Using a set of keywords (e.g., books, tutorials, course) we will filter project descriptions that are likely unrelated to software projects and will manually validate the projects captured by our heuristics. 

\begin{table}[]
    \centering
    \caption{Descriptive Statistics of the 887 candidate projects for our study.}
    \label{tab:project_stats}
    \begin{tabular}{l|r r r r}
    \toprule
        \textbf{Stats} & \textbf{Mean} & \textbf{Median} & \textbf{Min} & \textbf{Max} \\
    \midrule
        \# stars          & 5981.1 & 2195.0 & 133 & 144252 \\
        \# commits        & 3404.7 & 799.5 & 24 & 720863 \\
        \# contributors   & 68.7 & 29.0 & 9 & 466 \\
        \# of Issues      & 791.9 & 240.5 & 10 & 49558 \\
         age (years)    & 4.2 & 4.0 & 1.0 & 13.2 \\
    \bottomrule
    \end{tabular}
\end{table}

\noindent
\textbf{Datasets.}
From the selected projects, we plan to use multiple sources of information to answer our core research questions: 
\begin{itemize}
    \item \textbf{Model metadata}: Each PTM model metadata will be extracted from their hubs in Hugging Face Hub~\cite{HuggingFace:ModelHub:online}, and PyTorch Hub~\cite{PyTorch:online}. We plan to use this information to answer RQ1. 
    
    \item \textbf{Project source code}: We rely on source code analysis to answer RQ2 and RQ4.    
    
    \item \textbf{Project source code history}: We plan to traverse each project history to extract longitudinal data to answer RQ3. 
    
    \item \textbf{Project issues}: GitHub issues will be used to answer RQ5.  
\end{itemize}

\subsection{\textbf{\rqi}}
\textbf{Motivation:}
Pre-trained models (PTMs) are becoming integral to modern software development, yet their key characteristics, such as licensing, task domains, size, and architecture, are often underexplored. Understanding these attributes is crucial for enabling informed decisions for developers. Licensing, for instance, impacts compliance with open-source standards, while
analyzing the size and architecture of PTMs provides insights into computational requirements and performance potential. 

\begin{quotebox}
\textbf{Goal.} Provide a detailed overview of the PTMs used in OSS projects, based on the following characteristics: model license, task domain, model size, and model architecture.
\end{quotebox}

\textbf{Model Metadata Collection:}
To explore the characteristics of PTMs, we will retrieve metadata from two widely used hosting platforms: Hugging Face~\cite{HuggingFace:ModelHub:online} and PyTorch Hub~\cite{PyTorch:online}. Using their public APIs~\footnote{\url{https://huggingface.co/docs/hub/en/api}}~\footnote{\url{https://pytorch.org/docs/stable/hub.html}}, we will collect information about:
\begin{itemize}
    \item \textbf{License:} Identifying the licensing terms of each PTM to evaluate compliance and compatibility with open-source standards.
    \item \textbf{Task Domains:} Categorizing PTMs by their primary use cases, such as natural language processing (NLP), computer vision (CV), or multimodal applications.
    \item \textbf{Size:} Measuring the model size (e.g., in MB or parameter count) to assess computational demands and deployment feasibility.
    \item \textbf{Architecture:} Documenting the model architectures (e.g., transformer, ResNet) to analyze trends in model design and application.
\end{itemize}

We will use descriptive statistics, including mean, median, mode, and frequency distributions, to characterize PTMs based on licensing, task domain, size, and architecture. These distributions will be visualized clearly using bar charts and histograms.

\subsection{\textbf{\rqii}}
\label{sub:rq2}

\textbf{Motivation:} Pre-trained models (PTMs) may play diverse roles in software projects, ranging from core functionality to testing utilities and illustrative examples. Understanding how these models are embedded in the project requires a comprehensive analysis of their call sites and integration patterns. Additionally, examining whether PTMs are loaded statically (with hard-coded model names) or dynamically (through configuration files or environment variables) offers insights into software design practices and configurability. 

\begin{quotebox}
    \textbf{Goal:} Understand 1) the roles that PTMs play in the software project and 2) the coding practices related to loading PTMs in the code.  
\end{quotebox}

\textbf{Call Site Identification:} The first step in understanding PTM usage is to identify the \textit{PTM call sites}. 
The PTM call site is the part of the code responsible for loading the PTMs, and functions as a starting point for our analysis. 
Each PTM can be identified by their signature~\cite{Jiang:10.1145/3643991.3644907}, which contains: 

\begin{itemize}
    \item \textbf{File name:} The specific source code file containing the PTM call site.
    \item \textbf{Python Library name}: The library used to access HuggingFace Hub API or Pytorch Hub API to load the model.
    
    \item \textbf{Loading function call}: The function used to load the PTM using its respective Python API. 

    \item \textbf{Function parameters:} The parameters that often determine the model name, model size, and customized parameters. 
\end{itemize}

We plan to use the list of PTM signatures provided by the PeaTMOSS dataset~\cite{Jiang:10.1145/3643991.3644907}, which includes signatures commonly associated with PTMs. 
Using static code analysis tools, we built a parser that scans codebases to detect PTM call sites and usage. Each identified call site will be recorded with its file name, function name, and line number, creating a detailed map of PTM usage across repositories.
The file where the PTM call site is identified is hereafter named \textit{PTM loading file}.

\textbf{Assessing PTM Importance:} To assess the importance of PTM in their respective projects, we plan to analyse the dependencies between project files. 
We will build a file dependency graph of the project, where each node represents a file, and the directed edges represent dependency between files. For example, if file A imports methods from file B, an edge A $\to$ B is expected in the graph.
We will then measure the page-rank centrality of the PTM loading file, compared to other files of the project~\cite{PageRank}.
The page-rank centrality is commonly used to rank software artifacts in an ecosystem~\cite{Mujahid:Centrality:2022,Wittern:DynamicsOfJavascript:2016}. We specifically use page-rank centrality because it assigns importance to files not only based on direct dependencies but also by considering the importance of dependent files. Thus, model-loading files inherit significant centrality scores from highly important execution sites, effectively capturing indirect usage importance.

\begin{itemize}
    \item \textbf{Core functionality}: Projects where PTM loading files exhibit higher centrality, and PTMs are extensively used to accomplish different functionalities. 
    \item \textbf{Periphery functionality}: Projects where PTMs loading files exhibit low centrality and PTMs are used only during specific use-cases, e.g., testing, edge case scenarios. 
    \item \textbf{Disconnected from core project}: Projects where PTM loading files are disconnected from the connected component of the project core. This indicates that PTMs are loaded but not used by the main project, e.g., used as proof of concept, or tutorial. 
\end{itemize}

To better understand the role of PTMs within projects, we aim to qualify a sample of PTM models based on their usage context. For example, a PTM that is disconnected from the project may indicate that it is used for tutorials or examples, with no direct impact on the project's core functionality. We plan to use an open-coding approach, allowing themes to emerge from the data. Two independent annotators will code the context of PTM use, and inter-rater agreement will be assessed using the Cohen-Kappa metric to ensure reliability.
For sampling, we will employ a purposeful sampling strategy~\cite{sampling}, selecting PTMs from different projects that fall into three predefined importance categories: core functionality, periphery functionality, and disconnected. The authors will analyze the context of PTM usage within each project, including source code structure, dependencies, and associated documentation, to systematically categorize PTM roles and integration patterns.

\textbf{Analyzing Loading Patterns:} 
During our preliminary analysis, we noticed that projects tend to use distinct strategies in loading their PTMs. 
We plan to do a stratified sampling across projects to select a sample of the PTM call sites, and will classify them into 1) static loading, where the PTM model name and parameters are defined statically in the PTM loading file; and 2) dynamic loading, where the PTM configuration is fetched dynamically during runtime, e.g., via configuration files (\texttt{config['model\_name']}) and environment variables (e.g., \texttt{os.getenv()}). To define our analysis of dynamic loading patterns, we will distinguish between model configuration parameters and other dynamically loaded elements, such as authentication tokens. While configuration parameters define model behavior, authentication tokens control access and do not impact model execution. This distinction will be integrated into our qualitative analysis, ensuring a clearer understanding of how PTMs are configured and used beyond simply identifying static or dynamic loading.

\subsection{\textbf{\rqiii}}

\textbf{Motivation.} 
There is a very experimental nature in using PTMs for software projects. 
New PTMs are constantly being trained, and developers have to constantly adapt their code if they want to make use of the latest and the greatest models. 
Understanding how pre-trained models (PTMs) evolve within software projects is crucial for evaluating their lifecycle, longevity, and maintenance requirements. By examining PTM call sites over time, we aim to investigate how long PTMs remain in use and how frequently they are modified or updated to fulfill the project needs. 

\begin{quotebox}
    \textbf{Goal:} Understand how long PTMs remain in the code and how often the OSS projects update PTMs.  
\end{quotebox}

\textbf{Tracking PTM Call Sites Over Time:}
To study the lifecycle and evolution of PTMs, we propose tracking their call sites in repositories over a defined time period. We clearly define a PTM call site by its file name, function name, library name, and function parameters, ensuring consistent identification across commits. We will perform our analysis commit-by-commit, spanning from July 1, 2022, to December 31, 2024, representing a consistent 30-month analysis period. This period selection is justified by the significant surge in PTM adoption starting mid-2022, coinciding with important developments such as the popularity of ChatGPT and Stable Diffusion~\cite{Jiang:10.1145/3643991.3644907}.

\textbf{Model Longevity}
We define model longevity as the period that the PTM remains in the code base. 
As the same PTM can be loaded in different parts of the project code base (e.g., different files), we identify the PTM by its call site signature function call and parameters, which determines the model name, version, and customized parameters from the API.
For this analysis, we plan to compute the longevity of each PTM in all projects during the period of analysis (July 2022 to December 2024). 
This is calculated by the timespan between the PTM first appearance in the codebase and its last recorded presence.
We plan to report this aggregated data using the Kaplan-Meier survival analysis method~\cite{Kaplan-meyer:SurivalAnalysis}. 
The survival analysis is a nonparametric statistic used to measure the survival function from lifetime data.
In our case, the method is quite fitting to report how long PTMs "survive" in the code, and will allow us to identify stable PTM call sites that persist over time, as well as those that are short-lived or frequently modified.

\textbf{Model Family.}
We operationally define a “model family” using the explicit parent-child relationships provided by the Hugging Face API~\footnote{\url{https://huggingface.co/docs/huggingface_hub/main/en/package_reference/hf_api##huggingface_hub.HfApi.model_info}}. Specifically, models directly or indirectly linked to the same root parent node are grouped into a single family, ensuring clarity and reproducibility. For PyTorch Hub, the process is simpler given its smaller collection of 52 models, enabling straightforward categorization based on each model’s README documentation.

\textbf{Model Update Frequency.}
Alongside analysing the model longevity in the OSS projects, we plan to characterize how frequently developers update their models. 
As PTMs lack semantic versioning~\cite{ajibode2024semanticversioningopenpretrained}, we define PTM update, any replacement of a PTM callsite with another PTM callsite of the same model family. 
Models can be updated to the latest version (Llama2 to Llama3-7B) or to another variant of the same model (e.g., Llama3 to Llama3-4bit-quantized).
To identify these changes, we will use the insights reported by Ajibode et al.~\cite{ajibode2024semanticversioningopenpretrained} in the Hugging Face Model Hub and Pytorch Model Hub~\cite{HuggingFace:ModelHub:online}, to group models of the same family (e.g., Llama models, Gemini models).
When a PTM call site is replaced by another PTM call site of the same model family, we record a PTM update. 
We will apply Kaplan-Meier survival analysis to quantify PTM longevity and use descriptive statistics (e.g., frequency distributions, mean longevity) to summarize model update patterns.

\subsection{\textbf{\rqiv}}
\textbf{Motivation:} Testing is fundamental to maintaining software quality, enabling robust changes, reducing maintenance costs, and ensuring reliability. For projects leveraging pre-trained models (PTMs), testing is particularly crucial because PTMs require careful validation to ensure the quality of the PTM predictions and generation, to prevent unexpected failures and prevent breaking changes from model updates. Despite its importance, testing is often overlooked due to its complexity and the perceived time investment required to create effective test cases. 

\begin{quotebox}
    \textbf{Goal}: Understand if and how developers test the embedded PTMs in their open source projects.
\end{quotebox}

\textbf{Testing Practices Detection:}
To focus on analysing the testing practices that pertain the use of PTM in OSS projects, we plan to select projects that use common Python testing frameworks:
    \begin{itemize}
        \item We will analyze the project dependencies to identify the presence of popular testing libraries such as \texttt{unittest}, \texttt{pytest}, \texttt{nose}, or other testing frameworks. 
        \item Dependencies will be extracted from package managers (e.g., requirements.txt, Pipfile) or directly from source code analysis (\texttt{import pytest}).
        \item The focus on selecting only projects that use common Python testing frameworks stems from the fact that we will need to execute the tests to identify the testing practices related to PTMs. 
    \end{itemize}

\textbf{Extracting Code Coverage Metrics:}
Upon finding projects that include a testing framework, we plan to execute the test cases of the projects using the Python Coverage tool~\cite{Coverage64:online}.
The Coverage tool is a library that dynamically traces the execution of test cases (independently of the framework) by monitoring the execution trace and mapping it back to the source code.
We anticipate challenges in executing tests automatically in a large set of projects. 
Projects in which the tests cannot be automatically executed (e.g., unresolved dependencies) will be excluded from the analysis. 
When then plan to verify if the test coverage covers any of the PTM call sites. 
Covering the PTM call site indicates that, at some point, at least a single test has executed the model loading function from the PTM call site, which may indicate some level of PTM testing. 
Projects that cover at least one PTM call site in their test coverage will be manually inspected in the next step.

\begin{figure}[h]
\centering
\includegraphics[width=0.5\textwidth]{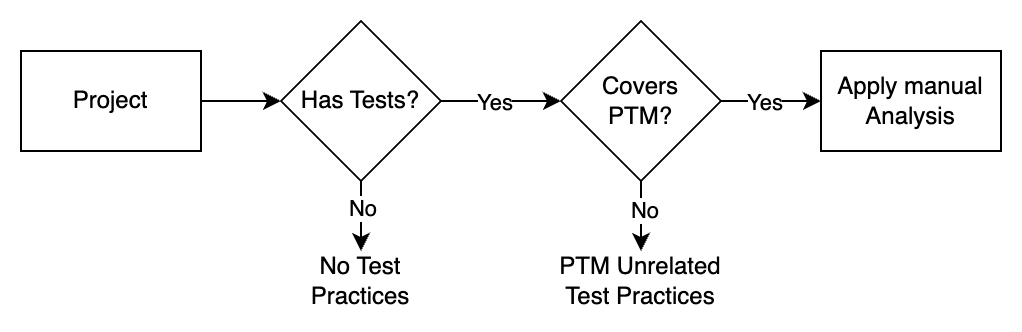}
\caption{Basic process of choosing projects for manual analysis}
\label{fig:methodology}
\end{figure}

We will employ a descriptive statistical analysis of test coverage data, explicitly summarizing test coverage distributions related to PTMs.

\textbf{Inspect PTM-Specific Testing:}
In this step, we plan to manually analyse the tests that cover PTM call sites from our selected projects.
When analyzing testing practices related to PTMs, our plan is to perform random sampling if the number of test sites is larger than 50, potentially aiming for a 90\% confidence level and a 5\% error margin. If the number of test sites is below 50, all of the sites will be analyzed.
The study will focus on determining:

\begin{itemize}
    \item \textbf{Types of PTM-related tests:} By analysing the project code and test suite, we plan to categorize the tests into test categories, as established by the literature and common software engineering practices~\cite{testing:hooda2015software} unit tests, integration tests, performance tests, stress tests.    
    
    \item \textbf{PTM testing practices:} In this step, we plan to evaluate qualitatively the coding practices, and the goal of the PTM testing. We will perform open coding~\cite{opencoding} independently by two annotators, followed by harmonization sessions to ensure conceptual consistency. Reliability and agreement between coders will be measured and reported using Cohen’s Kappa. We plan to identify recurring patterns of testing practices in Python coding. For example, is the PTM output ever asserted in software testing? Do developers encode the non-deterministic nature of commonly used PTMs (e.g., LLMs) in their testing?
    
\end{itemize}

\subsection{\textbf{PTM Issue Analysis}}
\textbf{Motivation:} Understanding the issues developers encounter when using pre-trained models (PTMs) in open-source software projects is essential for informing practitioners and improving PTM integration practices. PTMs introduce unique challenges, e.g., versioning conflicts, dependency mismatches, and performance inconsistencies. 

\begin{quotebox}
    \textbf{Goal:} Report on the most frequently identified themes in PTM-related issues reported in open source projects. 
\end{quotebox}

\textbf{Filtering PTM-Related Issues:} 
To investigate PTM-related issues, we will apply a snowballing approach~\cite{snowballing}, beginning with an initial list of keywords to identify a preliminary set of relevant issues. These keywords are derived from multiple sources, including PTM signatures identified in RQ1 (e.g., library names like transformers or torchvision, model names, and specific function calls), external platforms (e.g., Hugging Face, PyTorch Hub), file names and paths containing PTM call sites, and common PTM usage terms (e.g., ‘inference’ or ‘pre-trained’). Since file names and paths are unique per project, we will customize the keyword list for each project. After analyzing the preliminary set of issues, we will perform one iteration to expand and refine our keyword list, which will then be used to retrieve the final set of PTM-related issues.

\textbf{Manual Inspection and Labeling:} Once the filtering is complete, we apply open coding to a stratified sample of the filtered issues, ensuring stratification by project~\cite{opencoding}. Stratified sampling ensures that the sample is representative of the diversity of PTM-related issues across repositories. Both authors plan to independently review the sampled issues, labeling and categorizing them based on their content. Categories are not predefined but emerge during this manual inspection through harmonization sessions, allowing flexibility to capture the nuances of PTM-related challenges. Examples of potential categories include versioning problems, dependency conflicts, inadequate documentation, performance regressions, and compatibility issues.
We plan to report the interrater reliability metric using the Kappa statistic~\cite{Kappa}, and the two annotators will discuss conflicts to reach a consensus for the final coding.

\textbf{Analyzing Frequency and Distribution of Issues:} After the manual labeling and categorization are complete, we use the results to analyze the frequency and distribution of different issue categories. We will use descriptive statistics to analyze the frequency and distribution of categorized issues, clearly illustrating trends and patterns across identified PTM-related issues. This analysis provides insights into the most common and critical challenges developers face when integrating and maintaining PTMs.

\section{Threats to Validity}
This section discusses the potential threats that may affect the validity of our empirical study plan.

\noindent
\textbf{PeaTMOSS-derived Dataset.} Our initial project dataset is derived primarily from the PeaTMOSS dataset. Hence, threats to the internal validity of the PeaTMOSS dataset could affect the reliability of our results. Similarly, projects that have started using PTMs after the PeaTMOSS dataset was collected (late 2023) but became widely popular will not be included in our study. It is important to note, however, that the 887 selected projects will be analyzed with their latest snapshot to date, and we will consider their evolution beyond the PeaTMOSS publication date.

\noindent
\textbf{Incomplete Metadata (RQ1).} Since the primary source of metadata is the Hugging Face and Pytorch model hub, the completeness and accuracy of their recorded metadata could impact the validity of the findings. This is a well-known problem, as documentation practices are not standardized and well-established in model management~\cite{Jiang:10.1145/3643991.3644907}.

\noindent
\textbf{Classification Subjectivity (RQ2, RQ5).} Whenever manual and qualitative analysis is performed, we risk biasing our results towards the subjective experience of the annotators. To mitigate this threat, we plan to conduct the coding independently and report the interrater reliability metric to help the reader understand the discrepancy level between annotators, which could gauge the level of subjectivity of our analyses. 

\noindent
\textbf{Challenges in signature tracking (RQ3).} The historical analysis requires the matching of PTM call sites throughout the project history. PTM Call sites that load the model dynamically (i.e., the model name and URL are not static in the code) pose a severe challenge for our tracking. If dynamic loading sites dominate our dataset, we anticipate that our model update analysis may become less representative of real update practices in projects.   

\noindent
\textbf{Test Coverage Assessment (RQ4).} As we rely on executing tests for finding projects that test PTM, we might miss projects that are complex to build or adopt non-standardized testing methodologies. We do believe, however, that our demographic of projects will tend to follow common practices, leading to a more standardized way to execute tests.

\section{Conclusion}
In conclusion, this study plan aims to understand the integration, maintenance, and testing of pre-trained models (PTMs) in open-source software projects. 
By analyzing the metadata, code-base, testing practices, and issues of 887 OSS Python projects, we strive to uncover the challenges and best practices associated with integrated PTMs in real software. 
Our findings can potentially shed light on the evolving role of PTMs in OSS development and offer actionable insights for developers and researchers to enhance the reliability and sustainability of these models.

\bibliographystyle{IEEEtran}
\bibliography{references}

\clearpage
\end{document}